\documentstyle[12pt,epsfig]{article}
\def\Journal#1#2#3#4{{#1} {\bf #2}, #3 (#4)}
\def\Pramana{\em Pramana}
\def\jphysg{\em J. Phys. G, Nucl. and Part. Phys.}

\def\NPB{{\em Nucl. Phys.} B}
\def\PLB{{\em Phys. Lett.}  B}

\def\PRL{\em Phys. Rev. Lett.}
\def\PRD{{\em Phys. Rev.} D}
\def\ZPC{{\em Z. Phys.} C}

\def\ben{\begin{subequations}}
\def\be{\begin{equation}}
\def\een{\end{subequations}}
\def\ee{\end{equation}}
\def\bea{\begin{eqnarray}}
\def\eea{\end{eqnarray}}
\def\qsq{\mbox{$Q^2$}}
\def\kt{\mbox{$k_T$}}
\def\xgam{\mbox{$x_\gamma$}}

\def\f2gam{\mbox{$ F_2^\gamma $}}
\def\pt{\mbox{$p_T$}}
\def\et{\mbox{$E_T$}}
\def\siggg{\mbox{$\sigma_{\gamma \gamma}^{\rm inel}$}}

\def\vecqgam{\mbox{$\vec q_{\gamma} (x_\gamma)$}}
\def\qgam{\mbox{$q_{\gamma} (x_\gamma,Q^2)$}}
\def\eplem{\mbox{$e^+e^-$}}
\def\egam{\mbox{$e \gamma$}}
\def\ggam{\mbox{$g_{\gamma} (x_\gamma,Q^2)$}}
\def\cgam{\mbox{$c_{\gamma} (x_\gamma,Q^2)$}}
\def\glp{\mbox{$g_{p} (x_p,Q^2)$}}
\def\gamp{\mbox{$\gamma P$}}
\def\ep{\mbox{$eP$}}
\def\gamgam{\mbox{$\gamma \gamma$}}
\def\xgamobs{\mbox{$x_\gamma^{\rm OBS}$}}
\def\ptmin{\mbox{$p_{Tmin}$}}
\def\Phad{\mbox{$P_{had}$}}     
\def\siginel{\mbox{$\sigma^{inel}_{\gamma \gamma}$}}

\begin{document}
\begin{flushright}
hep-ph/yymmnn\\
IISc-CTS/7/98
\end{flushright}
\begin{center}
{\large{\bf Resolved Photon Processes}
\footnote{Plenary talk presented at Workshop on High Energy 
Physics Phenomenology, V, January 12-25, 1998, Pune, India.}}
\\[1cm]
R.M. Godbole \footnote{On leave of absence from
Dept. of Physics, Univ. of Bombay, India} \\
Center for Theoretical Studies, Indian Institute of Science, 
Bangalore, India.\\[0.5cm]
\end{center}

\begin{abstract}
After giving a very brief introduction to the resolved photon
processes, I will summarise the latest experimental information
from HERA, on resolved photon contribution to  large \pt\
jet production as well as to direct photon production. I will point
out the interesting role that resolved photon processes can play
in increasing our understanding of the dynamics of the Quarkonium
production. I will then discuss the newer information on the parton
content of virtual photons as well as the \kt\
distribution of the partons in the photon. I will end
by giving predictions of an eikonalised minijet model for   
\siggg\ which crucially uses the experimental measurement of the
abovementioned \kt\ distribution and comparing them with data.
\end{abstract}
\section* {Introduction}
The study of resolved photon processes~\cite{review} i.e., photon 
induced processes where the partons in the photon take part in the 
`hard' scattering process,  is useful to improve our understanding
of the interactions of high energy photons with other hadrons as well as
with each other. In addition, these processes  serve as an extra 
laboratory to study aspects of perturbative and nonperturbative QCD. 
The status of the study of the `resolved' processes at HERA and 
TRISTAN/LEP,  in relation to the  measurements of \f2gam\ at the \eplem\ 
colliders like PETRA/PEP/TRISTAN/LEP, is similar to the study of large 
\pt\ jet production at ISR which confirmed the parton structure of 
proton which had been revealed in 
the DIS experiments at SLAC. The first experimental verification of the 
`resolved' processes came at TRISTAN \cite{amy}. The higher energy 
and statistics at HERA \cite{herares} have since then provided a lot 
of interesting information  about photons and their interactions, 
in the study of these processes~\cite{maria}.  In this talk 
I summarise the current
experimental information on \f2gam\ and highlight interesting features
of the new analysis of jet production in resolved processes at HERA. I
discuss the role that the study of resolved contributions can play in
clarifying our understanding of the dynamics of the charmonium
production. I will then proceed to discuss new issues such as
resolved processes with photons of nonvanishing virtuality as well as 
the experimental determination of the transverse momentum (\kt)
distribution of partons in the photon. Lastly I will talk about
predictions for \siggg\ of an eikonalised minijet model, which uses 
the above measurement. 

\section{Photonic parton densities}
As is well known the photon structure function \f2gam\
describes the `strong' interactions that the $\gamma$ developes 
through  $\gamma \rightarrow q \bar q$ vertex.  A large amount
of data on \f2gam\ has been accumulated in the `hard' $e \gamma$
DIS scattering in the single tag \eplem\ experiments since the first
measurements of \f2gam\ by PLUTO \cite{pluto}.  The predictions of
perturbative QCD have two basic features:
\begin{itemize}
\item \f2gam\ rises linearly with  $\log \qsq $ and the scaling
violations are of an entrirely different nature as compared to the
case of the proton structure function. 
\item The parton densities in a photon \vecqgam\ peak at large \xgam\
unlike the parton densities in a proton.
\end{itemize}
\begin{figure}[htb]
\leavevmode
\begin{center}
%\vspace{6cm}
\mbox{\epsfig{file=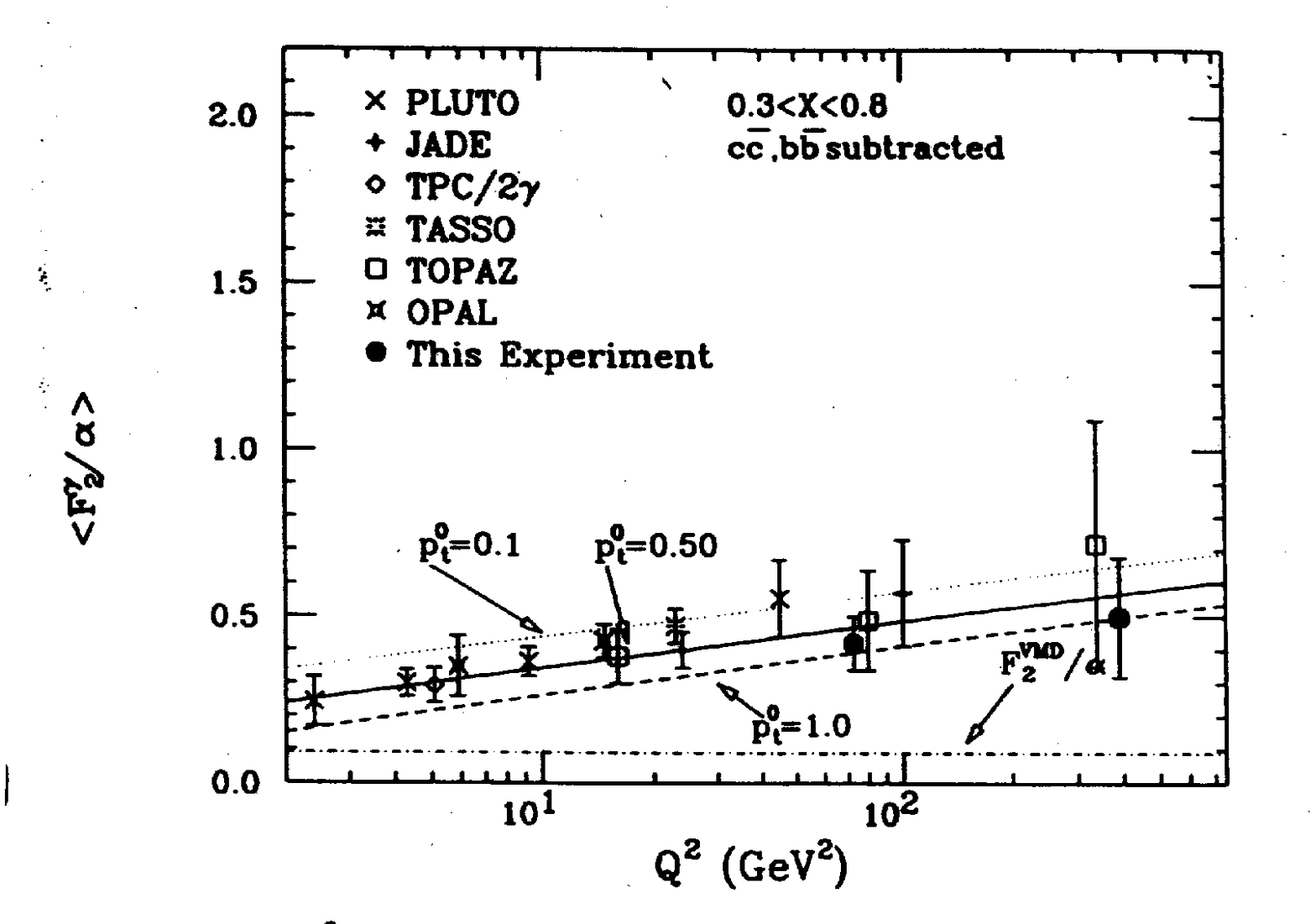,height=4cm}}
\end{center}
\caption{$x$ integrated \f2gam\ data as a function of \qsq}
\label{fig1} 
\end{figure}
Both these features have been experimentally verified. 
Fig. \ref{fig1}, taken from Ref.~\cite{amy2}, shows the expected 
linear  rise of integrated \f2gam\ (a factor of $\alpha_{\rm em}$ 
has been taken out)  with $\log \qsq$. A few more points can be made.
\begin{enumerate}
\item  The measurements of \f2gam\ , along with certain theoretical 
assumptions about the charm quark contribution, yield directly the
\qgam\ for light quarks over a wide range of $\xgam$ : 
$0.1\; < \; \xgam \; <  \; 1.0$, and  
upto $\qsq = 200\; {\rm GeV}^2$.

\item The gluon distribution \ggam\ is obtained from the DIS measurements 
only indirectly, using QCD evolution equations.

\item The charm quark distribution in photon, is correlated with the gluon
distribution \ggam\ and is not at all well determined. LEP2 offers a good
chance for its determination. 

\item As opposed to 1984, when there  was only one  parametrisation~\cite{dg}
for \vecqgam , obtained using fits to the early DIS data,
now at least twenty different parametrisations~\cite{parm} (inlcuding LO 
and NLO fits) are available. These are not just different fits but 
involve different assumptions about the nonperturbative part of the 
\f2gam\ and hence an accurate determination of \vecqgam\ will test 
these assumptions as well.  
\end{enumerate}
As a result of 1) above the \qgam\ in all the parametrisations are 
very similar. It is not so for \ggam. This can be seen from 
\begin{figure}[htb]
\leavevmode
\begin{center}
%\vspace{6cm}
\mbox{\epsfig{file=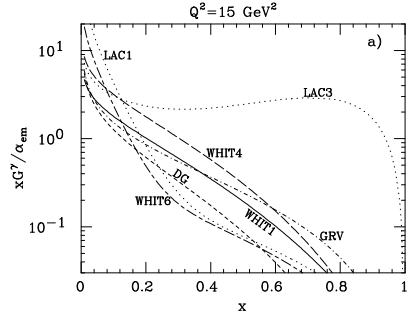,height=4cm}}
\end{center}
\caption{Gluon distributions in various parametrisations}
\label{fig2} 
\end{figure}
Fig.~\ref{fig2}. Hence, it is important to be able to obtain information
about \ggam\ from alternative sources.  Production of high \pt\ jets in 
effective \gamp\ and \gamgam\ collisions, which can be studied 
in photoproduction experiments at $eP$ collider (HERA) and in no-tag 
two-photon experiments in \eplem\ colliders (TRISTAN,LEP), was found to be
dominated by the `resolved' contributions \cite{herprd,tristanac} and hence
a good place  to study the gluon density in the photon.

\section{Jet production in resolved processes}
Although the original experimental observation~\cite{amy} of the resolved 
contribution  was
in jet production in \eplem\ collisions, the more recent experimental and
theoretical developement has happened in the study of jet production in
resolved processes in  \ep\ collisions. The
\begin{figure}[htb]
\leavevmode
\begin{center}
%\vspace{6cm}
\mbox{\epsfig{file=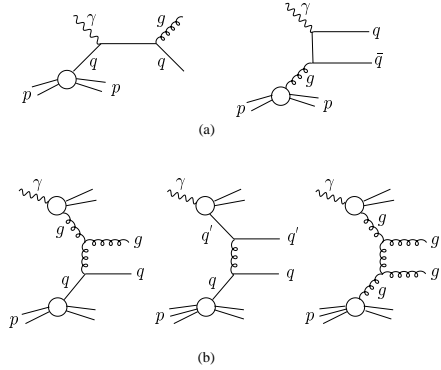,height=5cm}}
\end{center}
\caption{`Direct' and  `resolved' contributions to jet production in \gamp\
(\ep) collisions.}
\label{fig3} 
\end{figure}
two diagrams in Fig.~\ref{fig3}, give the `direct' and `resolved' contributions
respectively. In the `direct' (`resolved') processes all 
(only a fraction) of the photon energy is available for the hard scattering 
process. Hence the  fraction  of the photon energy carried by the jets,
\xgam , is $ 1 (< 1) $ for these two processes. This 
can separate the two processes if \xgam\ can be reconstructed  experimentally.
Note here that a neat separation between the two in terms of the 
diagrams is possible  {\it only} at the LO level. At NLO one 
has to be more careful in the separation 
between the two in the theoretical calculations. However, operationally
the difference between the two is clear:  only for the resolved processes
does one have a remnant jet activity in the direction of the photon (i.e.
the electron). This, along with the other two original 
predictions~\cite{herprd} , viz., 
that the `hard' resolved jets will be closer to the proton direction  than 
for the `direct' process and  that the former will dominate upto
large values of \pt , were qualitatively confirmed by the earliest 
measuements. 
\begin{figure}[htb]
\leavevmode
\begin{center}
%\vspace{6cm}
\mbox{\epsfig{file=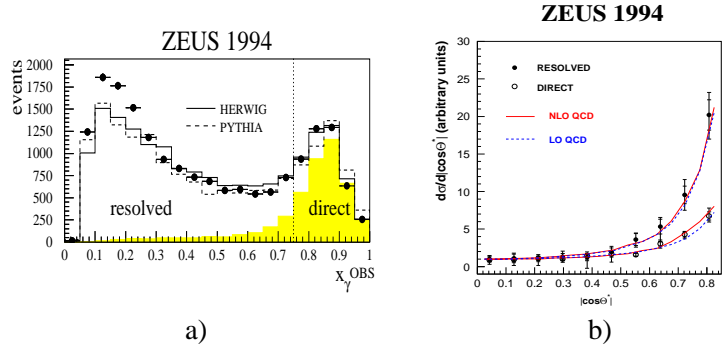,height=5cm}}
\end{center}
\caption{Experimental separation between  `direct' and `resolved'
contributions to inclusive jet production at HERA}
\label{fig4} 
\end{figure}
The dijet angular distributions expected in the LO direct(resolved) 
contribution are $|1-cos \theta^*|^{-1(-2)}$ corresponding to the 
spin 1/2(1) exchange in the  subprocess responsible for the 
jet production. The \xgamobs\ in Fig.~\ref{fig4}(a), taken from 
Ref.~\cite{dijets}, is essentially 
the fraction of the photon's momentum that appears in the high-\pt\
jets. As is seen from Fig.~\ref{fig4}(b),  the angular distributions
obtained by separating between  the resolved and direct processes 
based on $\xgamobs\ < \; 0.75$ and $\xgamobs\ > \; 0.75$ 
respectively, show this difference quite clearly.

As a matter of fact  early study of the differential distributions 
of the inclusive jet cross-sections, showed that photonic parton densities
as given by the LAC3 parametrisation, which was already disfaovured by
the \eplem\ jet data, was quite strongly ruled out,  despite  the 
uncertainties in the analysis which we shall mention next.

Since the early analysis, newer developements have been theoretical NLO
calculations~\cite{nlojets} and the incorporation of these in the 
analysis of the newer 1994/1995 HERA data~\cite{maria}. 
Even using LO analysis some general observations can be made. It is seen
\cite{review} that the dominance of the inclusive jet spectrum by 
the resolved contributions is somewhat reduced from the naive  
expectations, by stricter jet cuts; however, sensitivity to the \ggam\ 
can be increased by  rapidity cuts. It is therefore indeed possible 
~\cite{review} to try and use the data on inclusive jet 
photoproduction and dijet photoproduction to determine the \vecqgam .
However, 
inspite of the clear {\it qualitative}  evidence for the resolved processes 
provided by the data, a detailed analysis had showed  certain 
discrepancies between the data and theoretical predictions at low \et\ 
and high $\eta$ of the jet. It was realised that part of this is 
caused by the   `underlying ' event which can also 
contribute to the transverse energy activity around a `hard' jet. 
In the `resolved' events the remnant photon jet causes changes in the
hadronic activity of the event as compared to the direct events. It 
was demonstrated~\cite{desy9818}  that choice of a cone angle of 0.7 rather
than the ususal 1.0, is more appropriate for photoproduction of jets in the
resolved processes.  
\begin{figure}[htb]
\leavevmode
\begin{center}
%\vspace{6cm}
\mbox{\epsfig{file=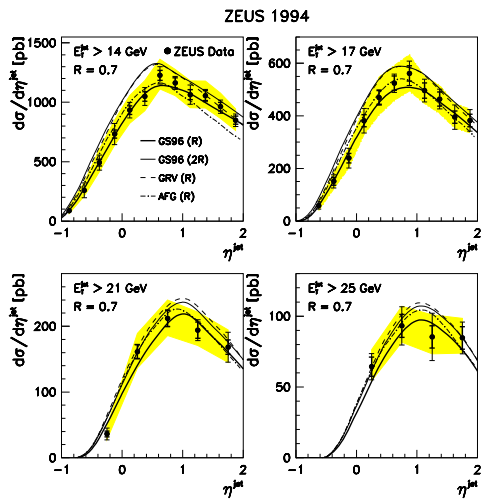,height=7cm}}
\end{center}
\caption{\et\ integrated rapidity distributions for inclusive jet
production.}
\label{fig5} 
\end{figure}
Fig.~\ref{fig5}, taken from \cite{desy9818}, shows that indeed the data 
are well described by the NLO calculations for all \et\ and $\eta$. 
In general, the data are better described by a MC which includes multiple
interactions (MI)~\cite{Butterworth}. The H1 data~\cite{h1pol1}
also show a direct correlation between the energy flow outside the 
jet and \xgamobs , again demonstrating the importance of an underlying 
event and need to understand it completely to analyse the dijet and
inclusive jet data. 

The dijet data have been used to extract an effective photon structure 
function
given by $ (q_\gamma(x_\gamma,Q^2) + \bar q_\gamma(x_\gamma,Q^2)) + 4/9 
g_\gamma(x_\gamma,Q^2))$, by H1 in anology to the $pp /p \bar p$ case.
\begin{figure}[htb]
\leavevmode
\begin{center}
%\vspace{6cm}
\mbox{\epsfig{file=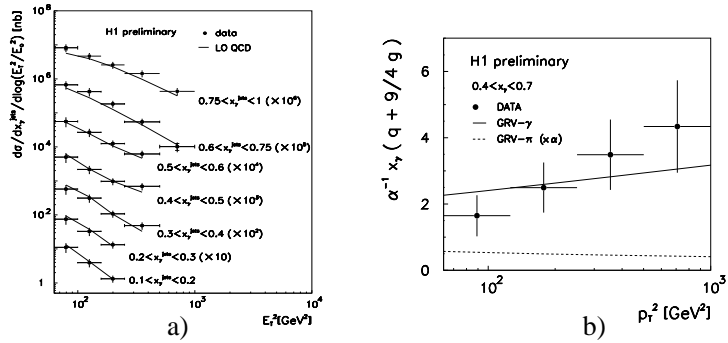,height=5cm}}
\end{center}
\caption{Dijet cross-sections in \ep\ collisions and extraction of an effective
parton distribution in the photon}
\label{fig6} 
\end{figure}
The dijet data are shown in Fig.~\ref{fig6}, taken from~\cite{h1pol2}. 
The effective parton densities in the photon shown in Fig.\ref{fig6}(b)
show  a striking linear rise with $\log \qsq$. However, the present 
data and analyses do not distinguish between the recent NLO
parametrisations like GRV,SAS and GS~\cite{parm}. 

The above discussion shows that the analysis of the inlcusive jet
cross-sections and dijet data has now come of age. A combined analysis
with the jet data from \eplem\ experiments, both TRISTAN and LEP-2,
along with the HERA data, are likely to increase our understanding
of the photonic parton densities.  Our increased knowledge of the
 protonic parton densities from HERA will play a very important role here.

\section{Direct photon and charmonium production}

The gluon distribution of a target is  probed effectively if one can
isolate hard processes where the gluon contributes dominantly. Both
at HERA and at the \eplem\ colliders, the direct photon 
and heavy quark production 
are subjects of experimental investigations to get  more information
about \ggam . The event rates for these are much smaller than 
the for jet production. However,
the extra information  they can provide makes   the study well worth it.
For direct photon studies, complete NLO calculation is available
\cite{diphnlo}. In the case of resolved processes there are additional 
technical issues involved as compared to the corresponding $pp / \bar p p$
case, due to the additional contributions coming from the fragmentation of
a parton into a photon. All of these issues have been satisfactorily 
addressed in the theoretical calculations.
The experimental investigations have just started~\cite{h1pol3} and the 
experiment has demonstrated the ability of  picking out the direct 
photon events. Unlike the case of jet production, here the 
resolved contribution ,though 
signficant, is not dominant. The current sample shows clearly only the 
existence  of the direct contribution. But analysis of more data certainly
holds promise.

Heavy quark and quarkonium production  is mostly dominated by the direct
contributions and the use of these to determine \vecqgam\  requires 
1)~a good knowledge of the \glp\ and 2) good set of cuts to separate 
the two from each other.  Actually, since heavy quark production 
at HERA is used as a probe for determination of \glp\ the latter 
are essential for that purpose as well.
The same kinematic feature which makes the jets in resolved processes come
out closer to the proton direction, can be used to isolate the two by using
simple rapidity cuts~\cite{review}. 

In case of quarkonium production
the second use of study of resolved processes is manifest. The dynamics of
quarkonium production  and the colour octet contribution to the production,
has been a subject of a lot of attention in the past few years. The size
of the colour octet matrix elements as extracted from the HERA data and from the
Tevatron data are in disagreement. One possible explanation is the 
apparent breakdown of some of the basic assumptions of the NRQCD 
calculations
in the region of large inelasticity $z$. As has been discussed 
elsewhere in the proceedings~\cite{michael}, it has been shown that the 
colour octet resolved contribution to $J/\psi$ production, dominates at
small inelasticity of $J/\psi$, where the uncertainty in the NRQCD 
calculations is small. Hence studying the $J/\psi$ production at small
$z$ at HERA will help pin down the  the colour octet matrix elements and
shed light on the discrepancy and might even rule out~\cite{eboli} 
the NRQCD model of $J/\psi$ production.
This is where one can see the use of resolved processes as a 
laboratory of QCD.

Study of inclusive charm production at HERA can also be used to obtain 
an effective LO measurement of the charm density \cgam ~\cite{sheffield}.
There also one can use effectively the same rapidity cuts as mentioned above
to separate the much more abundant direct contribution. 

Thus the field of better determination of gluon and charm quark density
in the photon using the abovementioned hard processes  is just opening up.
Theoretically the gluon and the charm densities in the photon are correlated.
The perturbative part of $F_2^{\gamma,c}$ is easily calculable. It is the
nonperturbative part, which can not be calculated, that  is closely correlated 
to the \ggam . The upcoming analysis of the two photon data
will offer a way to measure the charm contribution to \f2gam\ and hence 
shed more light on the issue.
 
\section{Internal \kt\  distribution of photonic partons and resolved
processes involving virtual photons.}
In a hadron like proton the internal transverse momentum distribution
of the partons is taken to be normally Gaussian. However, in the case
of photon since the parton content can in principle be traced back to
a hard $\gamma q \bar q$ vertex, the transverse momentum distribution
is expected to be different. It is expected  to be $f(\kt) \propto 
{1 \over (k_T^2  + k_0^2)}$ \cite{manuel,liame}.  ZEUS performed 
a determination of this distribution by essentially measuring the 
transverse momentum of the remnant jet (i.e the jet in the direction
of the electron) in a jet event. This measures the \kt\ distribution of
the parton which participated in the hard scattering process. The measurement
by ZEUS~\cite{zeuskt} gave a distribution in ageeement with the above
expectation and yielded a measurement of $k_0 = 0.66 \pm 0.22$ GeV,
This is indeed a preliminary measurement. Inclusion of such intrinsic
\kt\ has also improved the agreement with data on dijet rates and transverse
momentum distribution in \egam\ scattering in the analysis by OPAL.
In the next section I will discuss an eikonalised model prediction for
\siggg\ which makes use of this measurement.

The parton densities in  virtual photon can be completely computed
in perturbative QCD when the virtuality of the photon ($P^2$) is much 
higher than $\Lambda_{\rm QCD}^2$. For quasi real photons 
$(Q^2 \simeq 0)$   the parton densities can 
not be calculated from first principles, but those are the ones that are
parametrised from data. Since, in most photon-induced processes, 
the contribution to
the resolved cross-sections from virtual photons will be significant only
for low virtuality, $P^2 \sim \Lambda_{QCD}^2$, it is essential  to 
have a  model for these~\cite{dgvirt,SASvirt,GRS}. A measurement of 
jet production with virtual photons with small but {\it nonvanishing} 
virtuality, is now avalaible from H1~\cite{h1plb} and ZEUS~\cite{zeusjer} 
both. The current results are well described by some of 
the available parametrisations~\cite{dgvirt,SASvirt}. 
However, feasibility of the measurement is demonstrated and we can expect
more results from the analyses.

\section{\kt\ distribution of photonic partons and  model for total
inelastic cross-sections for \gamgam\ and \gamp\ processes}

The resolved contribution to hard processes, like jet production (say),
increases strongly with the energy of the photons. As a  matter of fact
it was observed~\cite{prlus} that the minijet cross-section,
\be
\sigma_{minijet}\equiv
\sigma (\gamma \gamma \rightarrow {\rm jets})\big|_{\ptmin}^{\sqrt s} \equiv
\int_{\ptmin} 
{d \sigma \over d \pt} (\gamgam \rightarrow {\rm jets})
\label{Sigjet}
\ee 
is completely dominated by the resolved contribution and rises like a power of
$s$, where $s$ is the square of the centre of mass energy. This
combined with the phenomenon of beamstrahlung can   cause 
very large backgrounds at the future linear colliders~\cite{messy}.
To assess this correctly one has to see how much of this rise gets
reflected in the rise of total inelastic cross-sections. This is also 
another example where investigations in issues involving
resolved processes, function as a theoretical laboratory. To that end here 
I will describe a model for the calculation of total inelastic cross-section
in  minijet model, trying to make a special effort to determine the
various parameters/inputs from the data involving resolved photons rather 
than as model assumptions.

The `minijet' cross-section has to be 
eikonalised so that unitarity is not violated. In general for photon 
induced processes, the inelastic cross-section obtained by eikonalisation
(and hence unitarisation) of the minijet cross-section is given by
\begin{equation}
\label{eikonal}
\sigma^{inel}_{ab} = P^{had}_{ab}\int d^2\vec{b}[1-e^{n(b,s)}]
\end{equation}
with the average number of collisions at a given impact
parameter $\vec{b}$ given by  
\begin{equation}
\label{av_n}
n(b,s)=A_{ab} (b) (\sigma^{soft}_{ab} + {{1}\over{P^{had}_{ab}}}
\sigma^{jet}_{ab})
\end{equation} 
where $P^{had}_{ab}$ is the probability that  the colliding particles
$a,b$ are both in a hadronic state,  
$A_{ab} (b)$ describes the transverse overlap of the  partons 
in the two projectiles  normalised to 1,
$\sigma^{soft}_{ab}$ is the non-perturbative part of the cross-section
while $\sigma^{jet}_{ab} $ is the hard part of the cross--section (of order
$\alpha_{\rm em} $ or $\alpha_{\rm em}^2$ for $\gamma p$ and 
$\gamma \gamma$ respectively). 
Notice that, in the above definitions, $\sigma_{soft}$
is a cross-section of hadronic size since the factor $P_{ab}^{had}$ has
already been factored out. 
We have,
\begin{equation}
\label{phad}
P_{\gamma p}^{had} = P_{\gamma}^{had}  \equiv P_{had} \ \ \ and \ \ \ 
 P_{\gamma \gamma}^{had} \approx   (P_{\gamma}^{had})^2.
\end{equation}
The overlap function $A_{ab} (b)$ is then,
\begin{equation}
\label{aob}
A_{ab}(b)={{1}\over{(2\pi)^2}}\int d^2\vec{q}{\cal F}_a(q) {\cal F}_b(q) 
e^{i\vec{q}\cdot \vec{b}}
\end{equation}
 where ${\cal F}$ is the Fourier transform of the b-distribution
of partons in the colliding particles. Normally, $A_{ab}$ is
 obtained using for ${\cal F}$ the electromagnetic form 
factors of the colliding hadrons. In general, for photons people have normally
used the form factor for a pion. We~\cite{liame} take a slightly different 
approach and calculate the `b- distribution' of the partons by taking the 
Fourier transform of the transverse momentum distribution of the partons, which
in the case of the photons is expected to be, at least for the perturbative 
part, 
\be
f(\kt) = {C \over (k_T^2 + k_{0}^2)}.
\ee
 As said in the earlier section $k_{0}$ has actually been measured 
by ZEUS to be 
$ 0.66 \pm 0.22$ GeV. It turns out that the form  of $A_{\gamma \gamma}$
with this transverse momentum ansatz and that for the pion form factor
ansatz, are the same, differing only in the value of the parameter 
$k_{0}$ which is $0.735$ GeV for the $\pi$ form factor case. Thus one can 
asses the effect of changing the ansatz for the $A_{ab}$ for photons by simply
changing the value of $k_{0}$.

For the soft part of the cross-section we use a parametrisation, 
\begin{equation}
\label{soft}
\sigma^{soft}_{\gamma p} =\sigma^0 +
{{A}\over{\sqrt{s}}}+{{B}\over{s}}
\end{equation}
we then calculate values for $\sigma^0, A$ and $B$ from a best fit 
\cite{thesis} to the low energy photoproduction data, 
starting with the Quark Parton Model (QPM) ansatz
$\sigma^0_{\gamma p}\approx {{2}\over{3}}\sigma^0_{pp}$ and the form 
factor ansatz for the $A_{\gamma p}$. The best fit value for \ptmin\ 
that we get is 2 GeV. It might be  possible to improve quality of the 
fit by using a energy dependent \Phad , but this needs to be 
investigted further.
The value of 2 GeV is also comparable to the value 1.6 GeV obtained~\cite{tjs}
from a fit to the description of minimum bias events in $pp/ \bar p p $ 
collisions.

For $\gamma \gamma$ collisions, we  repeat the QPM suggestion and propose
\begin{equation}
\sigma^{soft}_{\gamma \gamma}={{2}\over{3}} \sigma^{soft}_{\gamma p}.
\end{equation}
We now apply the  criteria and parameter set used in
 $\gamma p$ collisions to the case of photon-photon collisions,
i.e. $P_{\gamma}^{had}=1/204$, $\ptmin = 2$ GeV, 
A(b) from  the transverse momentum ansatz with the value $k_{0}=0.66$ 
GeV. The results of our calculation are shown 
\begin{figure}[hbt]
\leavevmode
\begin{center}
\mbox{\epsfig{file=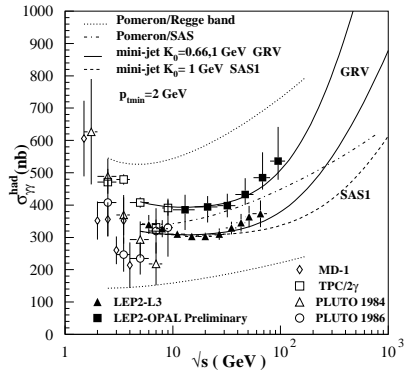,height=4cm}}
\end{center}
\caption{Total inelastic photon-photon cross-section in the eikonalised
 mini-jet model with  $\ptmin = 2\ GeV$,
 compared with data and Regge/Pomeron parametrization 
(see \protect~\cite{liame} for details). The two lower mini-jet 
curves correspond to $k_{0}=1$ GeV 
 with GRV and SAS1 densities. The highest one is for GRV densities and 
 $k_{0}=0.66$ GeV.}
\label{Gamgam}
\end{figure}
Fig.~\ref{Gamgam}. 
The highest of the two full  lines corresponds exactly to the
 same parameter set used in the photoproduction case
and appears to be in good agreement with the preliminary results from the
OPAL~\cite{OPAL} Collaboration, whereas the L3 results~\cite{LEPL3}, 
everything else being the same, would favour a higher $k_{0}$ value.

The following can be noticed here from the  newer data and 
model calculation:
\begin{itemize}
\item The data on \siginel\ rise faster with the energy than the predictions of
the Regge-Pomeron ideas and  the rise is consistent with predictions of the 
eikonalised minijet models where the parameters are fixed by fits to the
photoproduction case.

\item  These calculations use also more uptodate photonic parton densities
and the parameters of the eikonalised minijet model are related here to measured
properties of photon induced reactions.
\end{itemize}

\section{Conclusions}

Thus  we see that the HERA data have really lived upto the
expectations~\cite{herprd} that the photoproduction data, 
dominated by the resolved  processes, would yield information
on the photon structure. The study of two photon processes 
at the \eplem\ collider LEP2, and further analysis of heavy 
flavour (closed and open) production, direct photon production 
at HERA is sure to give more  information on the parton content
of the photon as well as on the way high energy photons interact
with each other and matter. In addition the resolved processes
are proving their potential as a laboratory for QCD.

\vspace{0.3cm}

{\bf Acknowledgements :} It  is a pleasure to thank M. Drees and 
G. Pancheri, for the enjoyable collaborative work, 
some of which has been discussed in this talk.


\begin{thebibliography}{98}
\bibitem{review}M. Drees and R.M. Godbole, \Journal{\Pramana} {41} {83} {1993};
\Journal{\jphysg}{21}{1559}{1995}.
\bibitem{amy} AMY Collab., R. Tanaka et al, \Journal{\PLB}{277}{215}{1992}.
\bibitem{herares} H1 Collab., T. Ahmed et al, \Journal{\PLB}{297}{206}{1992};
ZEUS Collab., M. Derrick et al, \Journal{\PLB}{297}{404}{1992}.
\bibitem{maria} For a recent and exhaustive review of the data on 
resolved photon processes see: M. Krawczyk, A. Zembrzuski and M. Staszel, 
{\bf hep-ph/9806291}.
\bibitem{pluto}PLUTO Collab., Ch. Berger et al, \Journal{\PLB}{107}{168}{1981};
\Journal{\PLB}{142}{111}{1984}.
\bibitem{amy2}AMY Collab., S.K. Sahu et al,\Journal{\PLB}{346}{208}{1995}.
\bibitem{dg}M. Drees and K. Grassie, \Journal{\ZPC}{28}{451}{1984}.
\bibitem{parm} For explanation of the shortforms for the different 
parametrisations used and the references for them, please see 
\protect\cite{review}.
\bibitem{herprd} M. Drees and R.M. Godbole, \Journal{\PRL}{61}{683}{1988};
\Journal{\PRD}{39}{169}{1989}.
\bibitem{tristanac}M. Drees and R.M. Godbole, \Journal{\NPB}{339}{355}{1990}.
\bibitem{dijets} ZEUS Collab., M. Derrick et al, \Journal{\PLB}{348}{665}
{1995}; Paper submitted to {\it ICHEP96, Warsaw}, 1996, {\bf pa 02-040}; 
\Journal{\PLB}{384}{401}{1996}.
\bibitem{nlojets} For a complete and uptodate list of references to NLO
calculations of inclusive jet and dijet cross-sections, see 
Ref.\protect~\cite{maria}. 
\bibitem{desy9818}ZEUS Collab., J.Breitweg et al, {\bf hep-ex/9802012},
(DESY-98-018); Paper submitted to the {\it HEP97 International 
Europhysics Conference on High Energy Physics}, 
Jerusalem, August 1997, {\bf N650}.
\bibitem{Butterworth} J.M. Butterworth, Talk presented at 
`Ringberg Workshop : New Trends in HERA physics', Tegernsee, May 1997,
{\bf hep-ex/9707001}.
\bibitem{h1pol1} H1 Collab., S. Aid et el, \Journal{\ZPC}{70}{17}{1996}.
\bibitem{h1pol2} H1 Collab., Paper submitted to {\it ICHEP 96,Warsaw}, 
{\bf pa 02-080}
\bibitem{diphnlo}  See for example L.E. Gordon and W. Vogelsang, 
\Journal{\PRD}{52}{58}{1995}. For a detailed list of references to the NLO
calculations see \protect\cite{review}.
\bibitem{h1pol3}H1 Collab., {\bf N265}, submitted to the same conference
as in \protect\cite{desy9818}.
\bibitem{michael} M. Kraemer, talk in these proceedings and references
therein.
\bibitem{eboli} O.J.P. Eboli, E.M. Greogores and F. Halzen, 
{\bf hep-ph/9802421}.
\bibitem{sheffield} M.Drees and R.M. Godbole, {\bf hep-ph/9505297},
{\it Proceedings of Photon-95}, Sheffield, April 1995, Published by World 
Scientific.
\bibitem{manuel} J. Field, E. Pietarinen and K. Kajantie, 
\Journal{\NPB}{171}{377}{1980};
M. Drees, {\it Proceedings of 23rd International 
Symposium on Multiparticle Dynamics}, Aspen, Colo., Sep. 1993. Eds. M.M. Block
and A.R. White. 
\bibitem{liame}
A. Corsetti, R.M. Godbole and G. Pancheri, in {\it Proccedings of 
PHOTON' 97},Eds. A. Buijs and F.C. Erne, Egmond aan Zee, May 1997,
World Scientific, {\bf hep-ph/9707360}; Also {\bf hep-ph/9807236}, 
To appear in Phys. Lett. B.
\bibitem{zeuskt} M. Derrick et al., ZEUS Coll., \Journal{\PLB}{354}{163}{1995}.
\bibitem{dgvirt} M. Drees and R.M. Godbole, \Journal{\PRD}{50}{1994}{3124}.
\bibitem{SASvirt}G. Schuler and T. Sjostrand, \Journal{\PLB}{376}{193}{1996}.
\bibitem{GRS}M. Gl\"uck, E. Reya and M. Strattman, \Journal{\PRD}{51}{3220}{1996}.
\bibitem{h1plb}H1 Collab., C. Adloff et al, \Journal{\PLB}{415}{418}{1997};
H. Rick(for H1 Collab.), {\em Proceedings of the PHOTON'97},
Eds. A. Buijs and F.C. Erne, Egmond aan Zee, May 1997,
World Scientific. 
\bibitem{zeusjer} ZEUS Collab., {\bf N 657}, submitted to the same conferecne
as in \protect\cite{desy9818}; also {\bf hep-ex/9508016}.
\bibitem{prlus} M. Drees and R.M. Godbole, \Journal{\PRL}{67}{1189}{1991}.
\bibitem{messy}M. Drees and R.M. Godbole, \Journal{\ZPC}{59}{591}{1993}.
\bibitem{thesis} A. Corsetti, September 1994 Laurea Thesis, University of Rome
La Sapienza.
\bibitem{tjs} T. Sjostrand and M. van Zijl, \Journal{\PRD}{36}{2019}{1987}.
\bibitem{OPAL} OPAL Collaboration :  F.W\"ackerle, to be published in the
Proceedings of   XXVII International Symposium on Multiparticle 
Dynamics, Frascati 8-12 September 1997, Conference
Suppl. Nucl. Phys. B, Eds. G. Capon et al. and
OPAL Physics Note 320, september 9, 1997. 
\bibitem{LEPL3} L3 Collaboration, \Journal{\PLB}{408}{450}{1997}.
\end{thebibliography}
\end{document}